\documentclass[a4paper,11pt]{article}
\usepackage{pos}

\title{The Fuzzy Onion: A proposal}

\author*[a,b]{S. Kov\'{a}\v{c}ik}
\author[a]{J. Tekel}

\affiliation[a]{Department of Theoretical Physics, Faculty of Mathematics, Physics and Informatics, \\
Comenius University in Bratislava, \\
Mlynsk\'a dolina, 842 48, Bratislava, Slovakia}
\affiliation[b]{Department of Theoretical Physics and Astrophysics, Faculty of Science, Masaryk University, \\
Brno, Czech Republic}

\emailAdd{samuel.kovacik@fmph.uniba.sk} 

\abstract{It is generally believed that the space has a nontrivial structure which is apparent on the order of the Planck length. There is a class of models of three-dimensional quantum spaces constructed using different mathematical tools. Also, there is another class of models with matrix descriptions of spaces of various dimensions and geometries with built-in momentum cut-off---these are called fuzzy spaces; the fuzzy sphere is a prominent example. In this paper, we describe how to connect various spheres together to foliate a three-dimensional space dubbed the fuzzy onion.}

\FullConference{%
 Corfu Summer Institute 2022 "School and Workshops on Elementary Particle Physics and Gravity",\\
 28 August - 1 October, 2022\\
 Corfu, Greece}

\begin{document}
\maketitle

\section{Introduction}

The fuzzy sphere is perhaps the most studied example of a fuzzy space and the simplest case of a quantized space \cite{Hoppe,Madore:1991bw,Kovacik:2013vbk,Galikova:2013zca}. In the common approach, one describes the field on the fuzzy sphere with a momentum cut-off using Hermitian matrices of a finite size. As the size of the matrices is increased, the cut-off is lifted. The machinery of the field theory on a sphere---and some other examples---can be recast in terms of matrix operators such as taking the trace instead of doing an integration. 

The crucial part of the model is the kinetic term of the action that is defined using the finite-size representation of the corresponding symmetry generators, for example, $SU(2)$ in the case of the sphere. Physics on fuzzy spaces comes with some advantages and disadvantages. Because of the natural cut-off, the theories do not suffer from various UV ill-effects. However, a typical feature of them is the UV/IR mixing which is a common aspect of nonlocal theories. As a result, the field theories in fuzzy spaces are known to possess phases that have no counterpart in ordinary scalar field theories even the infinite matrix size limit which is expected to recover the ordinary space. It is plausible that the standard formulation of the fuzzy theories' actions lacks higher-derivative terms that would hold those effects under control. Despite those difficulties, fuzzy spaces are of considerable interest. Firstly, they appear in various physical scenarios \cite{Steinacker:2011ix,Gubser:2000cd, Szabo:2001kg, Bagger:2007vi,Skenderis:2008qn,Baez:1998he} and, secondly, they are straightforward to study numerically \cite{Kovacik:2018thy,Ydri:2014rea, Panero:2006bx,GarciaFlores:2009hf}. 

Unfortunately, the known matrix models do not cover the case of the space we are interested in---that is the three-dimensional Euclidean space $\textbf{R}^3$. This is purely a problem on the matrix side, other formulations of $\textbf{R}^3$ are known and well-examined \cite{Vitale:2012dz, Kovacik:2013vbk,Galikova:2013zca}. Our goal is to formulate a matrix analog of those theories, to define a matrix formulation of the fuzzy $\textbf{R}^3$. The first idea is straightforward, one needs to glue together fuzzy spheres of different sizes. However, fuzzy spheres of different radii are described using matrices of different sizes and to take, for example, the radial derivative one needs to subtract matrices of incompatible sizes. We will describe here how to overcome this issue and formulate the fuzzy onion model using a set of concentric fuzzy spheres of increasing radius. 

\section{Scalar field on fuzzy spheres}

There are many ways how to think about the fuzzy sphere. Let us introduce it here in a way that allows us to extend the construction to a sequence of concentric fuzzy spheres. Scalar fields on an ordinary sphere can be expanded into spherical harmonics:

\begin{equation}
 f(\theta,\varphi) = \sum_{lm} c_{lm} Y_{lm} (\theta,\varphi),
\end{equation}
where $l=0,1,2,...$ and $m = - l, l+1 ,...,l$. The spherical harmonics are defined as eigenfunctions of the angular momentum operators that follow the $su(2)$ relations $[\hat{L}_i,\hat{L}_j]=i \varepsilon_{ijk} \hat{L}_k$:

\begin{eqnarray}
 \hat{{\cal{L}}} \ Y_{lm} &=& \hat{L}_i \hat{L}_i \ Y_{lm} = l (l+1) Y_{lm} , \\
 \hat{L}_3 \ Y_{lm}&=& m\ Y_{lm} .
\end{eqnarray}

There are of course also finite representations of these relations where $\hat{L}_i^{(N)}$ is an adjoined representation expressed in terms of $N \times N$ matrices. In this case also $Y_{lm}^{(N)}$ is a matrix of the same size and the angular momentum $l$ has an upper limit of $l=N-1$. 

Those two representations can be mapped onto each other if we impose the same limit also on the infinite-dimensional representation. Such truncated spectrum can only approximate the $\delta$-distribution. In other words, spatial resolution is limited and is fully recovered only in the infinite-$N$ limit. The relation between the size of the matrix, the size of the sphere, and the scale of space quantumness captured by the constant $\lambda$ is

\begin{equation}
r \sim N \lambda.
\end{equation}
Exact relation is a matter of convenience and can differ in various situations. It is now straightforward to construct a theory of a scalar field that exists on a series of disconnected fuzzy spheres of increasing radius. If we keep the constant of noncommutativity fixed across them we need to increase the corresponding matrix size to have increasing radii. Matrix with $N=1$ describes the innermost sphere, the $N=2$ matrix a layer above it and so on. We can put all of the into a single a block-diagonal matrix 

\begin{equation} \label{psi}
 \Psi = \begin{pmatrix}
\Phi^{(1)} & & &\\
 & \Phi^{(2)} & &\\
 & & \ddots &\\
 & & & \Phi^{(N_m)}
\end{pmatrix} 
\end{equation}
where $\Phi^{(N)}$ is a Hermitian matrix of size $N$. For a matrix to be interpretable as a fuzzy sphere we need to specify the Laplace operator. Its angular part is trivial

\begin{equation} \label{L}
 {\cal{L}} \Psi = \begin{pmatrix}
 \hat{{\cal{L}}}^{(1)}\Phi^{(1)} & & &\\
 & \hat{{\cal{L}}}^{(2)}\Phi^{(2)} & &\\
 & & \ddots &\\
 & & & \hat{{\cal{L}}}^{(N_m)}\Phi^{(N_m)}.
\end{pmatrix} ,
\end{equation}
where $ \hat{{\cal{L}}}^{(i)}$ is the Laplace operator acting of the fuzzy sphere of size $i$. 

\section{Scalar field on the fuzzy onion}

So far we have done little to nothing---only gathered a set of fields on $N_m$ fuzzy spheres and placed them into a single matrix $\Psi$. To have a proper field theory we need to add the radial part of the Laplace operator. To do so we need to define the radial derivative which is difficult as we need to define the radial derivative, $\partial_r \Phi^{(i)} \propto \Phi^{(i+1)} - \Phi^{(i-1)}$ which, as written here, is ill-defined due to subtraction of matrices of different size.

To connect consecutive spheres to obtain a three-dimensional theory, we need to define the following map

\begin{eqnarray}
 {\cal{U}} \Phi^{(i)} &=& \Phi^{(i+1)} \in {\cal{H}}(N+1), \\
 {\cal{D}} \Phi^{(i)} &=& \Phi^{(i-1)} \in {\cal{H}}(N-1), 
\end{eqnarray}
where $\Phi^{(i)}\in {\cal{H}}(N)$ and ${\cal{H}}$ is the set of Hermitian matrices of a given size. This map has to involve some kind of information loss as there is a different number of degrees of freedom on each of the layers due to different matrix sizes. The difference is in the modes of the highest angular momentum as they cannot be matched and it is therefore natural to remove or add those. 

This can be done in a straightforward way. To move one layer up, first, expand the matrix into spherical harmonics, then take the coefficients of the expansion $c_{lm}^{(N)}$ and define $c_{lm}^{(N+1)} =c_{lm}^{(N)}$ for $l \le N-1$ and $c^{(N+1)}_{l m} = 0$ for $l=N$. That means, mapping the coefficients to their corresponding counterparts and setting the rest to zero. To go one layer down, do the opposite: map the coefficients that can be mapped and remove the rest. As expressed together we have

\begin{eqnarray} \nonumber
 \Phi^{(N)} &=& \sum \limits_{l=1}^{N-1} \sum \limits_{m=-l}^l c^{(N)}_{lm} Y_{lm}^{(N)} \\ \nonumber
 &&\\ \nonumber
 &{\cal{D}} \uparrow \hspace{0.5cm} {\cal{U}} \downarrow &\\ \nonumber
 \Phi^{(N+1)} &=& \sum \limits_{l=1}^{N-1} \sum \limits_{m=-l}^l c^{(N+1)}_{lm} Y_{lm}^{(N+1)} \\ \nonumber
 && \\ \nonumber
 &\text{ where }& \\ \nonumber
 c_{l,m}^{(N)} &=& c_{l,m}^{(N+1)} \text{ for: } l\le N-1
 \\ \nonumber
 c_{N,m}^{(N+1)} &=& 0. 
\end{eqnarray} 
We can now define the first and the second-order derivative as \footnote{Note that here $\left(\partial_r\right)^2 \neq \partial^2_r$ as we find each of the definitions optimal---as an approximation---on its own.}

\begin{equation} \label{dr}
 \partial_r \Phi^{(N)} = \frac{{\cal{D}}\phi^{(N+1)} - {{\cal{U}}}\phi^{(N-1)} }{2\lambda},
\end{equation}
and 
\begin{equation} \label{ddr}
 \partial^2_r \Phi^{(N)} = \frac{{\cal{D}}\phi^{(N+1)} -2\phi^{(N)} + {{\cal{U}}}\phi^{(N-1)} }{\lambda^2}.
\end{equation}

Using equations \eqref{L}, \eqref{dr} and \eqref{ddr}, we can now define the full Laplace operator acting on the field $\Psi$ that lives on the fuzzy onion, which is formed using a set of concentric fuzzy spheres of increasing radii

\begin{equation}
 \Delta \Psi = \left(\frac{1}{r^2} \frac{\partial}{\partial_r} \left( r^2 \frac{\partial}{\partial r} \right) - \frac{{\cal{L}}^2}{ r^2} \right) \Psi
\end{equation}
where we have set $\hbar = 1$ and $r$ acts in a trivial way on each of the layers as $r\ \Phi^{(N)} = N \lambda \ \Phi^{(N)}$. 

Now we have a complete Laplace operator that contains the information about the underlying space. We call the space the field $\Psi$ exists on \textit{the fuzzy onion.}

\section{Potential term}

We have defined the kinetic part of the action of a scalar field living in a space foliated with concentric fuzzy spheres that we call the fuzzy onion for obvious reasons. It is a three-dimensional quantum space possessing rotational symmetry that resembles the construction in \cite{Kovacik:2013vbk,Galikova:2013zca, Vitale:2012dz}. There is a subtlety involved: the quantumness takes a different form in the angular and radial directions. While on each fuzzy spheres, it is realized as a momentum cut-off with no distinct structure---meaning there are no precedent points on the spheres, in the radial direction is the quantumness realized as a discrete, that is lattice, structure in which the half-line $r \in \textbf{R}_0^+$ is replaced by a set $r/ \lambda \in \textbf{N}$. So far we took the largest matrix to be of size $N=N_m$ so our space describes a (quantum) ball. By taking the limit $N_m \rightarrow \infty$ we can cover the entire three-dimensional space.

Another way the difference between radial and angular directions manifests itself is in the form of the potential. On a single fuzzy sphere the potential term is can be defined using a polynomial

\begin{equation} \label{fuzzy sphere potential}
V(\phi) = \mbox{Tr}\ P(\Phi),
\end{equation}
where the case of $P(\Phi) = b \Phi^2 + c \Phi^4$ is a prominent example well-studied in the literature \cite{Kovacik:2018thy,Ydri:2014rea, Panero:2006bx,GarciaFlores:2009hf,Tekel:2015zga}. A straightforward generalisation of this would be to take

\begin{equation}
V(\Psi) = \mbox{Tr}\ P(\Psi),
\end{equation}
so, for example, we can have 

\begin{equation} \label{pot}
V(\Psi) =\mbox{Tr} \begin{pmatrix}
b (\Phi^{(1)})^2 + c (\Phi^{(1)})^4 & &\\
 & \ddots &\\
 & & b (\Phi^{(N_m)})^2 + c (\Phi^{(N_m)})^4
\end{pmatrix} = \sum_{j=1}^{N_m} V(\phi^{(j)}).
\end{equation}
To put it in words, the total potential energy is the sum of the potential energies of all of the individual fuzzy spheres. An important feature of quantum spaces is the nonlocality. What does it mean in this context? A field on a fuzzy sphere cannot be limited to an area smaller than some elementary patch proportional to $\lambda^2$. That means that the value of the potential energy at some point also influences the value at other points in its vicinity. \footnote{Of course, the notion of a point is ill-defined here. We can speak of the north pole on the sphere for example, but what is meant by that is a blurred point-like density centered around this point. This fuzziness is the cause of the nonlocality of this construction.}

In \eqref{pot}, the nonlocality is felt across each of the individual fuzzy spheres but not between them. That means that if we produce a field excitation as localized as possible around the north pole of a given fuzzy sphere, it contributes to the potential energy close to that point on the same fuzzy sphere but not to the potential energy of the fuzzy sphere below and above it. The difference stems from the different forms of quantumness in the angular and radial directions.

There is a way how to solve it, how to connect various fuzzy spheres with quartic potential terms and that is to define it as 

\begin{equation}
V(\Psi) = b \mbox{Tr } \Psi^2 + c \left( \mbox{Tr } \Psi \right)^2.
\end{equation}

With this, the field on one layer feels the field on every other as the second term contains multi-trace contributions of the form $ \mbox{Tr }\ (\phi^{(i)})^2 \mbox{Tr } (\phi^{(j)})^2$ with $i \neq j$. This approach is the other extreme, in the first case all spheres were isolated, and here they are equally connected. Perhaps the best way would be a compromise where each layer effectively interacts only with its neighbors, that is to have contributions of the form $\mbox{Tr } (\phi^{(i)})^2 \mbox{Tr } (\phi^{(i+1)})^2$. Another option is to define a smeared value of a field in the radial direction of the form

\begin{equation}
{\cal{S}} \phi^{(n)} = \frac{\phi^{(n)}+ \sum \limits_{i} \alpha_i \left({\cal{U}}^i \phi^{(n-i)} + {\cal{D}}^i \phi^{(n+i)}\right)}{1 + \sum \limits_{i} \alpha_i},
\end{equation}
for example with $\alpha_1 = \frac{1}{2}, \alpha_{2+}=0$:
\begin{equation}
{\cal{S}} \phi^{(n)} = \frac{ \phi^{(n)} + \frac{1}{2} {\cal{D}}\phi^{(n+1)} + \frac{1}{2} {\cal{U}}\phi^{(n-1)}}{2}.
\end{equation}
Then we can take 
\begin{equation}
V(\Psi) = \sum \limits_{j=1}^{N_m} V \left({\cal{S}} \phi^{(j)}\right),
\end{equation}
where $V$ is defined in \eqref{fuzzy sphere potential} as an ordinary fuzzy-sphere potential but we are now using fields that have been smeared across various layers. 

\section{Conclusion}

In this paper, we have proposed how to connect a (potentially infinite) set of concentric fuzzy spheres of increasing vacua and the same spatial resolution limit to produce a three dimension quantum space that resembles previous theoretical constructions. An important aspect of this model is that the fields are encoded in a single Hermitian matrix of size $N_m \left(N_m+1\right)/2$ of a specific form. We call this model the fuzzy onion.

We have defined a procedure that maps between consecutive layers which allowed us to define the radial part of the Laplace operator and also a smearing procedure across various layers. There are various ways of defining the potential term and it would be interesting to study the differences in their behavior. 

A great feature of fuzzy space models expressed in terms of matrices is the accessibility of numerical simulations which offered great insight. There were various formulations of three-dimensional quantum spaces that were well-suited for analytical treatment. Our goal was to extend the fuzzy sphere model to cover the entire three-dimensional space while keeping the construction as close to it as possible to be able to have a connection to the large volume of research work done on it. 

The most straightforward application of the fuzzy onion model is to study the phase diagram of a quartic scalar field theory in three-dimensional space using the Hamiltonian Monte Carlo method. Another option is to study the behavior of three-dimensional objects, preferably having rotational symmetry; for example, a stellar core-collapse or normal modes of neutron stars but also heat dissipation in granular matter. 

Doing Monte Carlo simulations is not the only method to study the fuzzy onion model. In principle, many models expressed in terms of differential equations with the Laplace operator can be defined. A prominent example is the Schr\"odinger equation expressed as a matrix equation. Its solutions in the fuzzy onion model could be compared with those obtained in \cite{Galikova:2013zca} to test the validity of our approach. We will report on those efforts shortly. 

\acknowledgments
This research was supported by VEGA 1/0703/20 and VEGA 1/0025/23 grants and MUNI Award for Science and Humanities funded by the Grant Agency of Masaryk University.

\end{document}